\documentstyle[preprint,prb,aps,epsf]{revtex}
\begin{document}
\draft
\preprint{Shot noise, Dec. 1999}

\title{Interface effects on the
shot noise in normal metal- $d$ -wave superconductor
Junctions }

\author{Y. Tanaka, T. Asai, N. Yoshida, J. Inoue}

\address{Department of Applied Physics, Nagoya University, Nagoya,
464-8603, Japan.}

\author{S. Kashiwaya}

\address{Electrotechnical Laboratory, Umezono, Tsukuba, Ibaraki
305-8568, Japan.}

\date{\today}
\maketitle
\begin{abstract}
The current fluctuations in normal metal / $d$-wave superconductor
junctions are studied for various orientation of the
crystal by taking account of the spatial variation
of the pair potentials.
Not only the zero-energy Andreev bound 
states (ZES) but also the non-zero energy
Andreev bound states influence on 
the properties of differential shot noise.
At the tunneling limit,
the noise power to current ratio at zero voltage
becomes 0, once the ZES are formed at the interface.
Under the presence of a subdominant $s$-wave component at the interface
which breaks time-reversal symmetry,
the ratio becomes  $4e$.
\end{abstract}
\par
\newpage
\narrowtext
~
The origin of shot noise is the current
fluctuations in transport due to the discreteness of
the charge carrier.
Shot noise measurements provide
important information on conduction processes
which can not be obtained from the usual
conductance measurements
\cite{Jong1,Buttiker}.
In the last few years, several novel features peculiar to
shot noise in mesoscopic
systems have been revealed.
In particular, the shot noise in normal metal-superconducting junction
\cite{Khlus,Jong2,Muzykan,Anantram} and
superconductor / insulator / superconductor junctions
\cite{averin1,Rodero} have been intensively studied.
It has been shown, through these works, that
the Andreev reflection and the charge transport by the Cooper pairs
have significant influence on the transport fluctuation
at low voltages.
However, most of previous theories  are constructed
on the conventional $s$-wave superconductors  and
the theory  for $d$-wave  superconductors
has not been presented.
\par
On the other hand, extensive experimental and theoretical
investigations
have revealed that the pair potentials of
high-$T_{\rm c}$ superconductors
are $d_{x^{2}-y^{2}}$-wave symmetry.
\cite{Scalapino}
One of the essential differences of
$d_{x^{2}-y^{2}}$-wave superconductors
from conventional $s$-wave superconductors
is that the phase of the pair potential
strongly depends on the wave vector.
For example,
the appearance of zero-bias conductance peak (ZBCP)
in tunneling conductance at a (110) surface
of $d_{x^{2}-y^{2}}$-wave superconductors
reflects the sign change of effective
pair potential through
the reflection of
quasiparticle at the surface.
\cite{Hu,Tanaka1}
Orientational and material dependences
of ZBCP of
high-$T_{c}$ superconductors have been experimentally studied
in several groups,
and the consistency between theory and experiments
has been checked in details
\cite{Kashi95,Kashi96,Alff,Wei,Ig}.
\par
At this stage, it is an interesting problem to clarify what is expected
in the shot noise in normal metal / insulator
/$d_{x^{2}-y^{2}}$-wave superconductor $(n/I/d)$ junction
under the presence of the ZBCP.
Recently, Zhu and Ting presented a theory of the
shot noise in $n/I/d$ junction \cite{Zhu}.
They found  a remarkable feature for $n/I/d$ junction :
when the angle between the normal to the interface $\alpha$
is $\pm \pi/4$,
the noise-to-current ratio is zero
at zero-bias voltage and quickly reaches a classical Shottky
value $2e$ at finite voltage.
This feature is completely discrepant from that for
conventional $s$-wave superconductor
($n/I/s$) junctions
where the ratio is $4e$ at zero voltage and  $2e$ at finite voltage.
This anomalous behavior is responsible for the formation of
the zero energy
Andreev bound  states (ZES) at the interface
of the $d$-wave superconductor.
Although Zhu and Ting theory (ZT theory) clarified
important aspects of the
shot noise under the presence of ZES,
there still remain several unresolved problems.
\par
One is the detailed orientational dependence of the shot noise.
In ZT theory, the vanishment of
the noise-to-current ratio $R(eV)$ at zero voltage $(eV=0)$
is shown only for  $\alpha=\pi/4$ ($0 \leq \alpha \leq \pi/4$). 
Since this property is related to the existence of ZBCP in
tunneling conductance,
we must check the value of $R(0)$ for
$0 < \alpha <\pi/4$ where ZBCP appears in the tunneling conductance.
In this paper, we  will show that in the tunneling limit, $i.e.$,
low transparency limit, $R(0)$ vanishes for $\alpha \neq 0$
and becomes $4e$ only for $\alpha=0$
where no ZES are expected.
%In the same limit, we find that $R(0)$ is $2e$
%for normal metal / insulator / $p_{x} \pm i p_{y}$ superconductor
%junction
%where  ZES are only formed by the quasiparticles injected
%perpendicular to the interface.
Moreover, it is shown that $R(0)$ is classified into three values,
$0$, $2e$ and $4e$,
corresponding to
the region of the Fermi surface contributing to the ZES.
\par
The other is the
influence of the spatial dependence of the pair potential
on the shot noise.
It is known that when the ZES are formed at the interface of $d$-wave
superconductor the pair potential is suppressed
near the interface \cite{Buchholtz,nagato}.
Consequently, not only the ZES but also the
non-zero energy Andreev bound states (NZES) are formed \cite{Barash}.
The influences of the NZES on the shot noise
are clarified.
We further study the situation where a
subdominant $s$-wave component
which breaks the time reversal symmetry is induced near the interface of
$d$-wave superconductor \cite{Kuboki,Fogel,Matsumoto,Covington}.
The subdominant $s$-wave component influences significantly on
$R(eV)$.
%In the tunneling limit,
%$R(0)$ recovers to be $4e$ due to the disappearance of ZES.
\par
The model examined here is  a two-dimensional $n/I/d$
junction within the quasiclassical formalism
\cite{Bruder,Millis}
where the pair potential has a spatial dependence
\begin{equation}
\label{A1}
\bar{\Delta}(x,\theta)
=
\left\{
\begin{array}{cc}
0,& (x \leq 0)\\
\bar{\Delta}_{R}(x,\theta), &
(x \geq 0)
\end{array}
\right.
\end{equation}
Here $\theta$ is the angle of quasiparticle trajectory
measured from the $x$ axis.
If we apply this formula to $d$-wave superconductors
including a subdominant $s$-wave component near the interface,
$\bar{\Delta}_{R}(x,\theta)$ is decomposed into
\begin{equation}
\label{A2}
\bar{\Delta}_{R}(x,\theta)
=\Delta_{d}(x)\cos[2(\theta - \alpha)]
+ \Delta_{s}(x)
\end{equation}
where $\alpha$ denotes the angle between the normal to the
interface and the $x$ axis of the crystal.
The insulator located between the normal metal and the  
superconductor is modeled by a $\delta$ function.
The magnitude
of the $\delta$-function denoted as $H$ determines the
transparency of the junction $\sigma_{N}$, with
$\sigma_{N}=\cos^{2}\theta / [Z^{2} + \cos^{2} \theta]$
and $Z=mH/\hbar^{2}k_{F}$.
The effective mass $m$ and Fermi momentum $k_{F}$ are assumed to be
constant throughout the junction.
The noise power to current ratio $R(eV)$,
the differential shot noise $S_{T}(eV)$,
and the tunneling conductance $\sigma_{S}(eV)$,
are given by \cite{Anantram,Zhu}
\begin{equation}
\label{A3}
R(eV)=
\frac{\int^{eV}_{0} dE S_{T}(E) }
{ \int^{eV}_{0} dE  \sigma_{S}(E) }
\end{equation}

\begin{equation}
\label{A4}
S_{T}(eV)=\frac{1}{2} \int^{\pi/2}_{-\pi/2} d\theta
\bar{S}(eV,\theta) \cos\theta,   \ \ \
\sigma_{S}(eV)=\frac{1}{2}\int^{\pi/2}_{-\pi/2} d\theta
\bar{\sigma}_{S}(eV,\theta) \cos\theta
\end{equation}

\begin{equation}
\label{A5}
\bar{S}(eV,\theta)=\frac{4e^{3}}{h}
[R_{a}(1-R_{a}) + R_{b}(1-R_{b}) + 2R_{a}R_{b}]
\end{equation}

\begin{equation}
\label{A5n}
\bar{\sigma}_{S}(E,\theta)=\frac{2e^{2}}{h} (1 + R_{a} -R_{b})
\end{equation}

The magnitude of the
Andreev and normal reflection
$R_{a}$ and $R_{b}$ are given by \cite{Andreev,Blonder,Zaitsev}
\begin{equation}
\label{A6}
R_{a}
= \frac{ \sigma_{N}^{2} \left|{\eta_{R,+}(0,\theta)}\right| ^{2} }
{ \left| {1 + (\sigma_{N} -1) \eta_{R,+}(0,\theta)
\eta_{R,-}(0,\theta)}
\right| ^{2} },
\end{equation}
and
\begin{equation}
\label{A7}
R_{b}
= \frac{ (1-\sigma_{N}) \left|1 -
\eta_{R,+}(0,\theta) \eta_{R,-}(0,\theta)
\right| ^{2} }
{ \left| {1 + (\sigma_{N} -1) \eta_{R,+}(0,\theta)
\eta_{R,-}(0,\theta)}
\right| ^{2} }.
\end{equation}
As a reference,  we also calculate the tunneling conductance
normalized by that in the normal state,

\begin{equation}
\label{A8}
\sigma_{T}(eV)=
\frac{\int^{eV}_{0} dE \sigma_{S}(E)}
{\frac{1}{2} \int^{\pi/2}_{-\pi/2} d\theta \sigma_{N} \cos\theta}. \
%\sigma_{R}(E,\theta)
%= \frac{ 1+ \sigma_{N} \left|{\eta_{R,+}(0,\theta)}\right| ^{2}
%+ (\sigma_{N}-1 )\left|  \eta_{R,+}(0,\theta)
%\eta_{R,-}(0,\theta)
%\right|^{2} }
%{ \left| {1 + (\sigma_{N} -1) \eta_{R,+}(0,\theta)
%\eta_{R,-}(0,\theta)}
%\right| ^{2} }.
\end{equation}
Note that the differential shot noise and
conductance spectrum are expressed only by
$\eta_{R,\pm}(x,\theta)$ just at the boundary $(x=0)$
where $\eta_{R,\pm}(x,\theta)$ obeys the following equations,
\begin{equation}
\label{A9}
\frac{d}{dx} \eta_{R,+}(x,\theta)=
\frac{ 1}{i \hbar {v}_{F}\cos\theta }
\left[ -\bar{\Delta}_{R}(x,\theta_{+})
\eta_{R,+}^{2}(x,\theta)  -\bar{\Delta}_{R}^{*}(x,\theta_{+})
+ 2 E \eta_{R,+}(x,\theta) \right],
\end{equation}
\begin{equation}
\label{A10}
\frac{d}{dx} \eta_{R,-}
(x,\theta)=
\frac{ 1}{i \hbar v_{F}\cos\theta }
\left[ -\bar{\Delta}_{R}^{*}(x,\theta_{-}) \eta_{R,-}^{2}
(x,\theta)  -\bar{\Delta}_{R}(x,\theta_{-})
+2E \eta_{R,-}(x,\theta) \right],
\end{equation}
with $v_{F}=k_{F}/m$, $\theta_{+}=\theta$ and $\theta_{-}=\pi-\theta$,
and $\bar{\Delta}_{R}(x,\theta_{+})$ $[\bar{\Delta}_{R}(x,\theta_{-})]$
is the effective pair potential felt by an electron [a hole] like
quasiparticle. 
The quasiparticle energy $E$ is measured from the Fermi energy. 
\par
The spatial dependence of the pair potentials are determined by the following
equations \cite{ashida}
\begin{equation}
\label{B1}
\Delta_{s}(x)=g_{s}k_{B}T\sum_{\omega_{n}}
\frac{1}{2\pi}\int^{\pi/2}_{-\pi/2} d\theta'
\{
[g_{R}(\theta',x)]_{12}-[g^{+}_{R}(\theta',x)]_{12}\}
\end{equation}

\begin{equation}
\label{B2}
\Delta_{d}(x)=g_{d}k_{B}T\sum_{\omega_{n}}
\frac{1}{2\pi}\int^{\pi/2}_{-\pi/2} d\theta'
\cos[2(\theta'-\alpha)]
\{
[g_{R}(\theta',x)]_{12}-[g^{+}_{R}(\theta',x)]_{12}\}
\end{equation}

\begin{equation}
\label{B21}
\lim_{x \rightarrow \infty} \Delta_{s}(x)= 0, \
\lim_{x \rightarrow \infty} \Delta_{d}(x)=\Delta_{0}
\end{equation}
with dimensionless inter-electron potential of the $s$-wave
$g_{s}$ and  $d$-wave $g_{d}$, respectively.
The quasiclassical Green's function $g_{R}(\theta,x)$ obeys
\cite{ashida}
\begin{equation}
\label{B3}
g_{R}(\theta,x)=U_{R}(\theta,x,0)g_{R}(\theta,0)
U_{R}^{-1}(\theta,x,0)
\end{equation}

\begin{equation}
\label{B4}
i\hbar v_{Fx}\frac{\partial}{\partial x}
U_{R}(\theta,x,0)
= -
\left(
\begin{array}{ll}
i\omega_{n} & \Delta_{R}(x,\theta_{+}) \\
-\Delta^{*}_{R}(x,\theta_{+}) & -i\omega_{n}
\end{array}
\right)
U_{R}(\theta,x,0), 
\end{equation}
with $\omega_{n}=2\pi k_{B}T(n+1/2)$ and $U_{R}(\theta,0,0)=1$. 
In the actual numerical calculations, $\eta_{R,\pm}(x,\theta)$
is calculated from Eqs. (\ref{A9}) to (\ref{A10}).
Since $g_{R}(\theta,0)$ is expressed by
$\eta_{R,\pm}(\theta,0)$, $g_{R}(\theta,x)$ is obtained
using Eqs. (\ref{B3}) to (\ref{B4}).
Subsequently, the spatial dependence of the pair potentials 
$\Delta_{d}(x)$ and
$\Delta_{s}(x)$ are calculated by Eqs. (\ref{B1}) to (\ref{B21}).
To get self-consistently determined pair potential,
this process is repeated until enough convergence is obtained.
\par
First let us consider the case where $\Delta_{s}(x)$ is not present.
It is known
for $\alpha \neq 0$, $\sigma_{T}(eV)$ has a zero bias conductance peak
(ZBCP) with small magnitude of $\sigma_{N}$ 
\cite{Tanaka1,Kashi95,Kashi96}.
It is expected that this property reflects on
$S_{T}(eV)$ for $\alpha \neq 0$.
In Fig. 1, $S_{T}(eV)$ is plotted for $Z=5$ with $\alpha=\pi/6$
[see curve $a$ in Fig. 1(a)].
As a reference,  similar calculations are performed based on the
non self-consistent pair potential where the spatial dependence of the
pair potential is chosen as $\Delta_{d}(x)=\Delta_{0}$
[see curve $b$ in Fig. 1(a)]).
As seen from  curves $a$ and $b$, $S_{T}(0)$ has a peak
around zero voltage originating from the
ZBCP in $\sigma_{T}(0)$ [see Fig. 1(b)].
As compared to curves $a$ to $b$,
both the height and width of the peak around zero voltage
of curve $a$
are small as compared to those in $b$,
since the degree of resonance at zero voltage is weakened
due to the reduction of $\Delta_{d}(x)$  at the interface,
Besides this property,  curves $a$
in %\sigma_{T}(eV)$
and $S_{T}(eV)$ ($\sigma_{T}(eV)$)
have second peak around
$eV \sim 0.45 \Delta_{0}$
due to the formation of NZES which can not be
expected in curves $b$. \par
In Fig. 2, we plot $R(eV)$ for $n/I/d$ junction
for sufficient low transparency case, $e.g.$, $Z=5$.
In the case of $\alpha=0$,
no ZES are formed at the interface, then the resulting
$R(eV)$ is $4e$ at zero voltage and is $2e$ at higher voltage
(curve $a$ in Fig. 2).
When  the value of $\alpha$ deviates from  $0$, $R(0)$ is zero and it
quickly reaches a classical Shottky
value $2e$, which is consistent with
other report \cite{Zhu} (curve $b$ and $c$).
This feature is explained based on the $\theta$ dependence of
$\bar{\sigma}_{S}(0,\theta)$ and $\bar{S}(0,\theta)$
as follows:
in general, $\bar{\sigma}_{S}(0,\theta)=\frac{4e^{2}}{h}R_{a}$
and $\bar{S}(0,\theta)=\frac{16e^{3}}{h}R_{a}(1-R_{a})$
are satisfied.
At the tunneling limit, $R_{a}$ is suppressed  with the increase of $Z$
unless the ZES  are formed at the interface.
When ZES are formed, $R_{a}=1$ is satisfied independent of $Z$.
Such a situation is realized for
$\pi/4-\alpha <\mid \theta \mid <\pi/4 + \alpha$.
For $\alpha=0$, no ZES are formed and
both the magnitude of $S_{T}(0)$ and
$\sigma_{S}(0)$ are reduced with the increase of $Z$,
while $R(0)$ remains to be constant.
On the other hand, for $\alpha \neq 0$ the magnitude of
$S_{T}(0)$ is reduced while $\sigma_{S}(0)$ remains finite
with the increase of $Z$, thus $R(0)$ becomes zero.
The important point is that  the origin of the vanishment of
$R(0)$ is responsible for the presence of ZES.
It is a universal property excepted for unconventional superconductor
junctions where finite region of the Fermi surface contributing on the
formation of the ZES.
\par
The critical situation
is realized in
$n/I/p_{x}+ip_{y}$ junction where ZES are formed only
by the  quasiparticles injected perpendicular to the interface
\cite{Zwicknagl}.
The spatial dependence of the pair potential
$\bar{\Delta}_{R}(x,\theta)$ is given by
%%%%%%%%%%%%%%%%%%%%%%%%%%%%%%%%
%  p-wave
%%%%%%%%%%%%%%%%%%%%%%%%%%%%%%%%
\begin{equation}
\label{C1}
\bar{\Delta}_{R}(x,\theta)
=  \Delta_{p1}(x)\cos \theta
+  i \Delta_{p2}(x) \sin \theta
\end{equation}
with
\begin{equation}
\label{C2}
\lim_{x \rightarrow \infty} \Delta_{p1}(x) =\Delta_{0}, \
\lim_{x \rightarrow \infty} \Delta_{p2}(x) =\Delta_{0}.
\end{equation}
$R(eV)$ is calculated based on the
self-consistently determined pair potentials,
$\Delta_{p1}(x)$ and $\Delta_{p2}(x)$.
As shown in curve $d$ in Fig. 2,
$R(eV)$  is nearly $2e$ independent of bias voltages.
The nature that $R(0)$ is neither $4e$ nor $0$
has never been predicted
in previous theories \cite{Anantram,Zhu}.
\par
%%%%%%%%%%%%%%%%%%%%%%%%%%%%%%%%%%%%%%%%%%%%
% d +is state
%%%%%%%%%%%%%%%%%%%%%%%%%%%%%%%%%%%%%%%%%%%%%%%
Under the presence of the ZBCP, since the quasiparticle
density of states at the zero energy near the interface are enhanced, 
a subdominant $s$-wave component of the pair potential $\Delta_{s}(x)$ 
can be induced near the interface, 
when a finite $s$-wave pairing interaction strength exists,
even though the bulk symmetry 
remains pure $d$-wave \cite{Kuboki,Fogel,Matsumoto,Covington}.
Since the phase difference of the
$d$-wave and $s$-wave components is not a multiple of $\pi$,
the mixed state breaks the time-reversal symmetry 
\cite{Fogel,Matsumoto}. 
In such a case, the ZBCP splits into two and
the amplitude of the splitting depends on
the magnitude of the induced $s$-wave component.
The corresponding  $R(eV)$ is plotted in Fig.3(a)
with $\alpha=\pi/4$ and $Z=5$ where the transition temperature
of $s$-wave component is chosen as $T_{s}=0.15T_{d}$ (dotted line)
and $T_{s}=0.3T_{d}$ (solid line).
The induced $s$-wave component near the interface influences crucially
on the $R(eV)$ at low voltages.
The remarkable feature is that $R(0)$ recovers to be $4e$ as in the
cases of $n/I/s$  junctions.
It is because that
with the inducement of $s$-wave component which breaks the
time reversal symmetry, the position of the Andreev bound state
shifts to $eV \sim \Delta_{s}(0)$, 
where $\sigma_{T}(eV)$ has a peak [see Fig. 3(b)].
Consequently, ZBCP shifts to this voltage
and $R(eV)$ has a dip structure.
Although $R(0)=4e$ is satisfied, the overall feature of
$R(eV)$ is completely
different from that in
$n/I/s$ junction, since $s$-wave component is induced
only near the interface
in the present case.
\par
In this paper,
the current fluctuations in normal metal / $d$-wave superconductor
junctions are studied for various orientation of the
junction by taking account of the spatial variation
of the pair potentials.
Not only the zero energy Andreev bound 
states but also the non-zero energy 
Andreev bound states show up in the line shape of $S_{T}(eV)$.
At the tunneling limit, we found universal property of
$R(0)$ for  two dimensional superconductors.
$R(0)$ is $0$, $2e$ and $4e$, 
corresponding to three cases where
the region of the Fermi surface contributing to the
ZES is i)finite region, ii)point and iii)none, respectively.
The present property gives useful information
on the identification for the
symmetry of the unconventional superconductors.
We hope the measurement of shot noise in unconventional superconductors
will be performed near future.
\par
It is known that ZES also influences significantly on
Josephson current in
$d$-wave superconductor / insulator / $d$-wave superconductor
($d/I/d$) junctions \cite{Tanaka2,Barash1}.
It is an interesting and future problem to study the
shot noise in $d/I/d$ junctions.
\par
%\vspace{1.0cm}
%This work was partially supported by the
%Core Research for Evolutional
%Science and Technology (CREST) of the Japan
%Science and Technology
%Corporation (JST) and a Grant-in-aid for Scientific Research
%from the Ministry of Education, Science, Sports and Culture. \par

%
\noindent
\newpage
\noindent
%%%%%%%%%%%%%%%%%%%%%%%%%%%%%%%
%  Figure
%%%%%%%%%%%%%%%%%%%%%%%%%%%%%%%
%%%%%%Figure1
\begin{figure}
\vspace{20pt}
%\figureheight{1cm}
\begin{center}
\leavevmode
 \epsfxsize=80mm
 \epsfbox{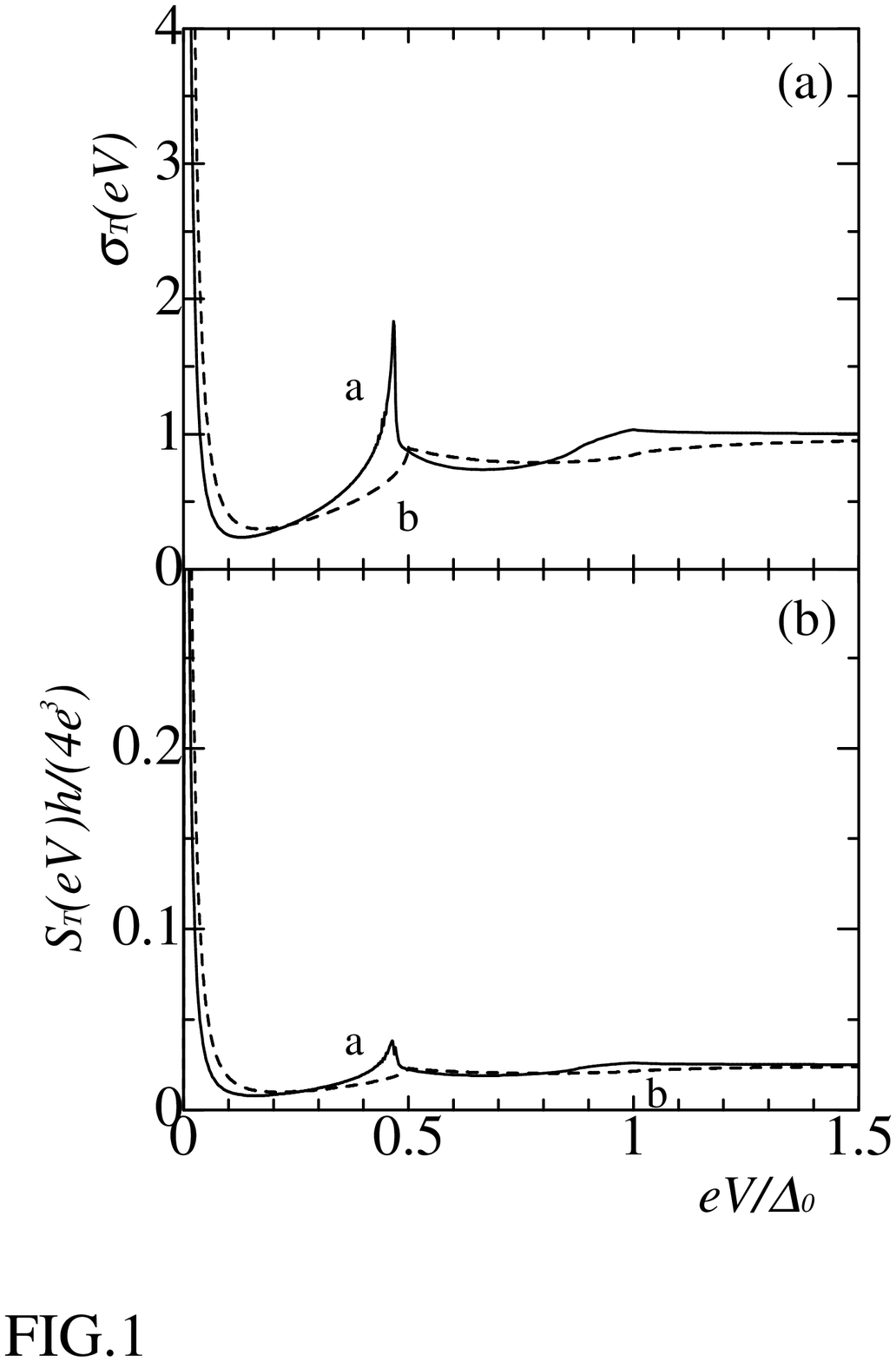}
\end{center}
\caption{$\sigma_{T}(eV)$ [Fig.1(a)] and $S_{T}(eV)$ [Fig.1(b)]
for $n/I/d$ junction with $Z=5$ and $\alpha=\pi/6$.
a: self-consistent calculation,
b: non self-consistent calculation. }
\label{fig:f1}
\end{figure}
%%%%Figure2
\begin{figure}
\vspace{20pt}
\begin{center}
\leavevmode
 \epsfxsize=80mm
 \epsfbox{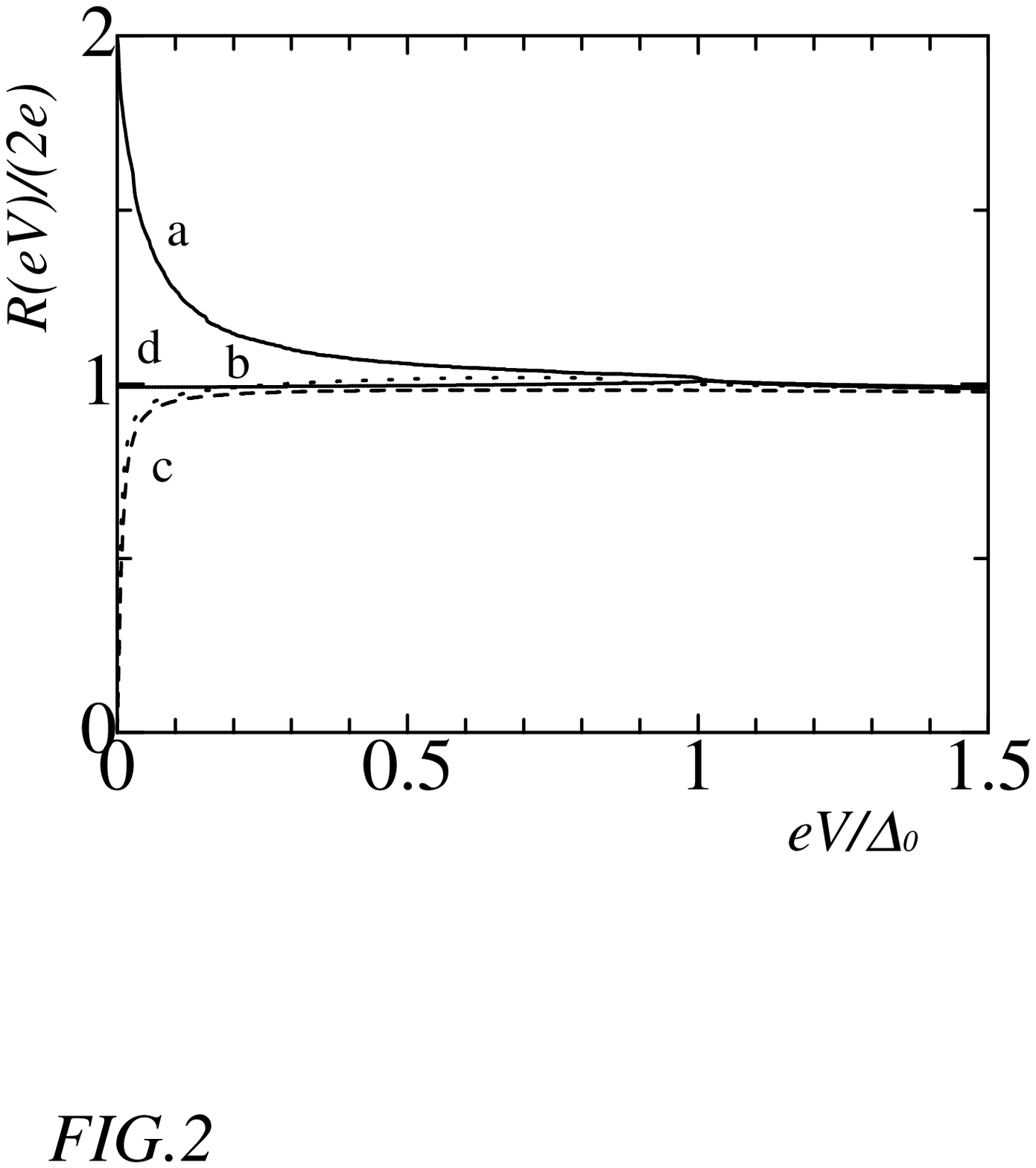}
\end{center}
\caption{$R(eV)$ is plotted with $Z=5$
a: $\alpha=0$
b: $\alpha=\pi/12$
c: $\alpha=\pi/4$
for $n/I/d$ junctions.
d: Similar plot for $n/I/p$ junction.}
\label{fig:f2}
\end{figure}
%%%%%
%%%%%%%%%Figure3
\begin{figure}
\vspace{20pt}
\begin{center}
\leavevmode
 \epsfxsize=80mm
 \epsfbox{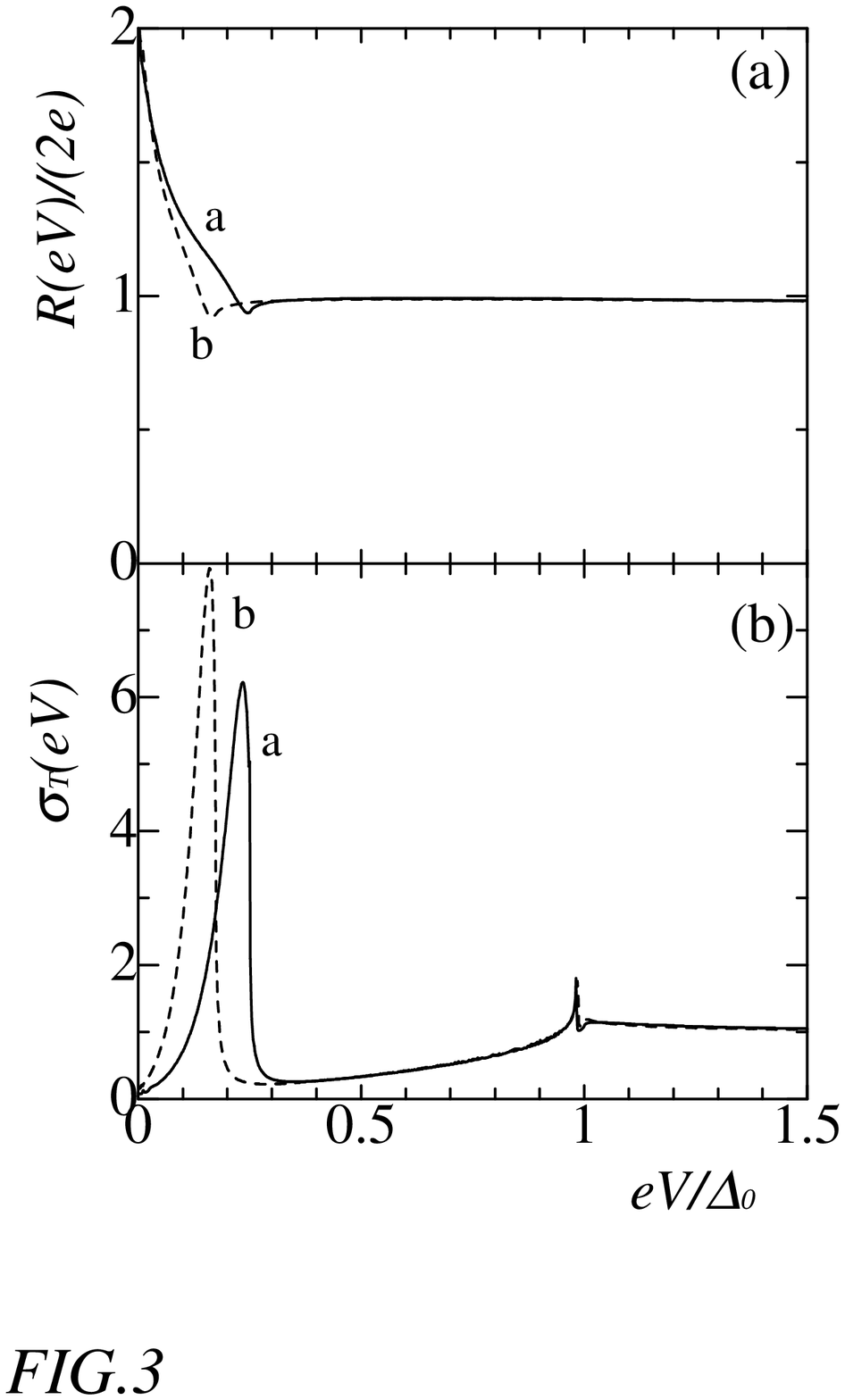}
\end{center}
\caption{$R(eV)$ is plotted for $Z=5$ and $\alpha=\pi/4$
where $s$-wave component is induced at the interface [Fig. 3(a)].
a: $T_{s}=0.15T_{d}$ 
and  b: $T_{s}=0.3T_{d}$.
The corresponding $\sigma_{T}(eV)$ is plotted in [Fig. 3(b)].}
\label{fig:f3}
\end{figure}
\end{document}